\documentclass[draft]{agujournal2019}
\usepackage{url}
\usepackage{soul}
\usepackage{lmodern}
\usepackage{gensymb}
\usepackage{xcolor}

\draftfalse

\journalname{Geophysical Research Letters}

\begin{document}

\title{Earth's mesosphere during possible encounters with massive interstellar clouds 2 and 7 million years ago}

\authors{Jesse A. Miller\affil{1,2}, Merav Opher\affil{1,2}, Maria Hatzaki\affil{3}, Kyriakoula Papachristopoulou\affil{3,4}, and Brian C. Thomas\affil{5}}

\affiliation{1}{Department of Astronomy, Boston University, Boston, MA 02215}
\affiliation{2}{SHIELD NASA DRIVE Center}
\affiliation{3}{Section of Geography and Climatology, Department of Geology and Geoenvironment, National and Kapodistrian University of Athens, GR-15784, Athens, Greece}
\affiliation{4}{Institute for Astronomy, Astrophysics, Space Applications and Remote Sensing, National Observatory of Athens (IAASARS/NOA), Athens, GR-15236, Greece}
\affiliation{5}{Department of Physics and Astronomy, Washburn University, Topeka, KS 66621}

\correspondingauthor{Jesse Miller}{jesmill@bu.edu}

\begin{keypoints}
\item As a result of colliding with interstellar clouds 2 and 7 million years ago, Earth's upper atmosphere received an abundance of hydrogen
\item By converting interstellar hydrogen to water in the lower thermosphere, thick noctilucent clouds would have formed
\item HO$_x$ compounds could have depleted mesospheric ozone by up to 99\%, though the total ozone column generally increases
\end{keypoints}

%% ------------------------------------------------------------------------ 

\begin{abstract}

Our solar system's path has recently been shown to potentially intersect dense interstellar clouds 2 and 7 million years ago: the Local Lynx of Cold Cloud and the edge of the Local Bubble. These clouds compressed the heliosphere, directly exposing Earth to the interstellar medium. Previous studies that examined climate effects of these encounters argued for an induced ice age due to the formation of global noctilucent clouds (NLCs).
Here, we revisit such studies with a modern 2D atmospheric chemistry model using  parameters of global heliospheric magnetohydrodynamic models as input. We show that NLCs remain confined to polar latitudes and short seasonal lifetimes during these dense cloud crossings lasting $\sim10^5$ years. Polar mesospheric ozone becomes significantly depleted, but the total ozone column broadly increases. Furthermore, we show that the densest NLCs lessen the amount of sunlight reaching the surface instantaneously by up to 7\% while halving outgoing longwave radiation.

\end{abstract}

\section*{Plain Language Summary}

As the Solar System moves through the interstellar medium, it encounters different astrophysical environments. By tracing back the path of the Sun, two possible crossings of dense interstellar clouds 2 and 7 million years ago have been identified. These clouds are dense enough to compress the solar wind to inside of Earth's orbit, exposing Earth's atmosphere to interstellar gas. Previous studies that explored terrestrial climate changes due to these event argued for a global cooling effect that could trigger an ice age.
In this study, we revisit this topic with a modern computational atmospheric chemistry model. We find that high-altitude water significantly enhances the density and coverage of noctilucent clouds (NLCs) near the mesopause. In contrast with previous studies, this effect is neither permanent nor global, though some denser NLCs may still block up to 7\% of sunlight from reaching Earth's surface. Furthermore, HO$_x$ compounds greatly deplete mesospheric ozone. We find for the first time that this mesospheric ozone decrease allows for a stratospheric ozone increase, resulting in an increase in the total ozone column. In order to assess the complete global climate response to these events, a more complete 3D model is required.

%%%%%%%%%%%%%%%%%%%%%%%%%%%%%%%%%%%%%%%%%%%%%%%%%%%%%%%%%%%%%%%%%%%%%%%%%%%
\section{Introduction}
\begin{figure}
\centering
\noindent\includegraphics[width=\textwidth]{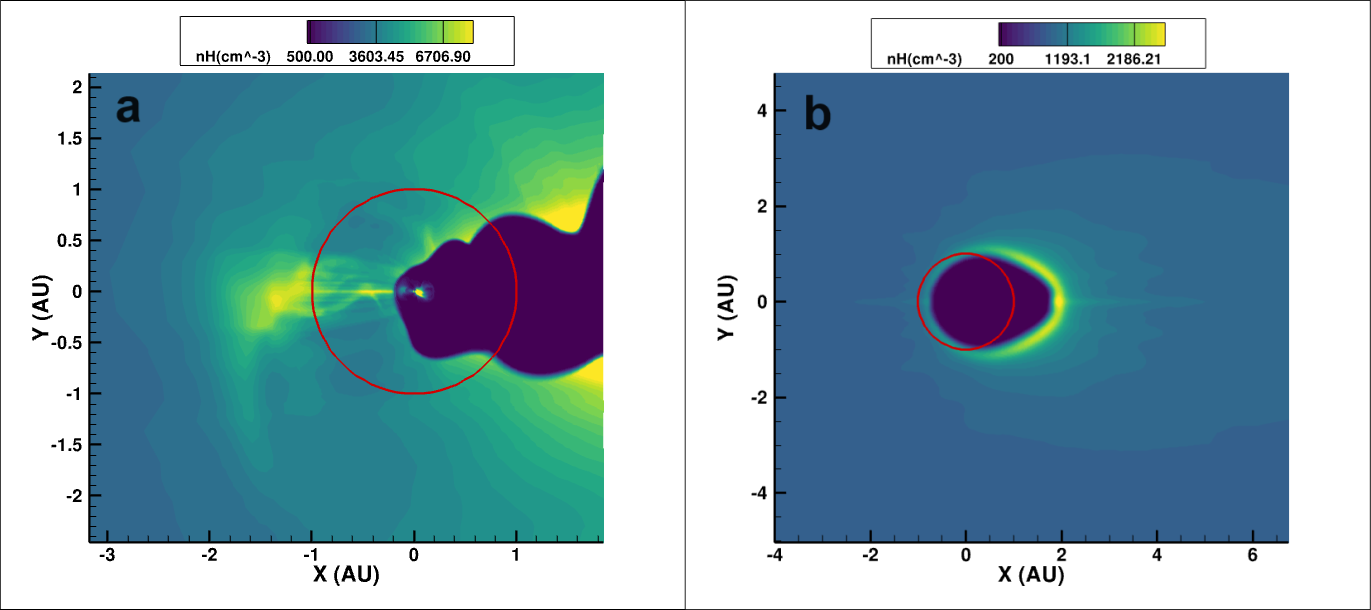}
\caption{Equatorial cuts from MHD simulations \cite{opher_possible_2024, opher_passage} showing the compression of the heliosphere by interstellar clouds with densities of a) 3000 cm$^{-3}$ and b) 900 cm$^{-3}$, where the color contours show neutral hydrogen density. These events may have occurred 2 and 7 Myr ago as our solar system collided with the LxCC and the edge of the Local Bubble, respectively. The red circle shows Earth's orbit. One can see that Earth experiences densities up to 6000 cm$^{-3}$ in case (a) and 1500 cm$^{-3}$ in case (b).}
\label{fig:heliosphere}
\end{figure}

The current environment outside the solar system is that of a partially ionized interstellar medium with a density of 0.1 cm$^{-3}$. The solar system moves at 18 pc/Myr along its path through the interstellar medium (ISM) and encounters many different interstellar environments. One of the most extreme is the passage through a dense interstellar cloud. By tracing back our solar system's path, \citeA{opher_possible_2024} recently found that we intersected with such a cloud 2 Myr ago, the Local Lynx of Cold Cloud (LxCC), which form the tail of the Local Ribbon of Cold Clouds (LRCC). This cloud had a density high enough ($\sim$3000 cm$^{-3}$) to compress the heliosphere to within Earth's orbit. Consequently, Earth was no longer shielded by the heliosphere, but ``descreened'' and exposed directly to the interstellar medium. Another similar descreening event may have happened when our solar system traversed the edge of the Local Bubble 6.8 Myr ago. Fig. \ref{fig:heliosphere} shows equatorial planes from MHD simulations of the heliosphere during these crossings, demonstrating the interstellar hydrogen density to which Earth was exposed \cite{opher_possible_2024, opher_passage}. The MHD simulations accounted for the velocity differences between the Sun and the LxCC and the expansion of the Local Bubble. For example, the velocity difference between the LxCC and the Sun is 18.5 km/s at large distances, but due to the Sun's gravitation potential, the velocities reach 40 km/s at Earth's orbit (not shown). For model conditions and details, see \citeA{opher_possible_2024, opher_passage}.
 We will take these values as an input to the climate model described below.

The two clouds described in these recent studies \cite{opher_possible_2024, opher_passage} have densities similar to those of molecular clouds. Various studies have examined how Earth's exposure to a giant molecular cloud (GMC) could affect our climate \cite{hoyle_effect_1939, mccrea_ice_1975, talbot_climatic_1976, begelman_can_1976, mckay_consequences_1978, zank_consequences_1999, yeghikyan_consequences_2003, yeghikyan_terrestrial_2004, pavlov_catastrophic_2005, pavlov_passing_2005}.
However, these studies were all justified by a statistical argument of how often our solar system intersects GMCs in the Galaxy. Such an event is expected to occur with a frequency of 1-10 Gyr$^{-1}$ \cite{smith_habitable_2009, talbot_climatic_1976}.

The LxCC has not been as thoroughly studied as the most popular member of the LRCC, the Local Leo Cold Cloud (LLCC). However, \citeA{opher_possible_2024} argue that since both the LxCC and LLCC are members of the LRCC \cite{haud_gaussian_2010},
the density and structure of the LxCC and LLCC should be similar. The LLCC was shown to have a relatively slender profile \cite{meyer_remarkable_2012}. Hence, we expect that the LxCC would be similarly thin. It is not considered a GMC, but it still achieves the high density required to compress the heliosphere to 1 au ($\sim$3000 cm$^{-3}$). Similarly, the walls of the Local Bubble have dense star-forming clouds that we may have crossed around 6.8 Myr ago \cite{zucker_star_2022}. \citeA{opher_passage} assume those clouds had a density of 900 cm$^{-3}$. In this paper, motivated by the recent connections between the heliosphere and nearby interstellar clouds, we examine atmospheric effects of two specific events, passing through the LxCC and the edge of the Local Bubble, within the past 10 Myr.

Previous studies that modeled climate effects of encounters with interstellar clouds with densities of 3000 cm$^{-3}$ at the top of the atmosphere \cite{mckay_consequences_1978, yeghikyan_terrestrial_2004} used simplified chemistry schemes. They predicted that there would be a long-term (essentially the whole duration of Earth's transversal of the GMC, assuming its extent of $\sim$1 pc takes $\sim$10$^5$ years) layer of global noctilucent clouds (NLCs). This NLC layer would induce a global cooling by blocking sunlight and cause an ice age. Here we revisit such calculations with a modern 2D atmospheric chemistry model and focus on the mesosphere.
In contrast to previous assertions, the majority of Earth's surface will not be under NLCs for the entire interstellar cloud crossing. NLCs maintain their seasonality, lasting no more than six weeks out of the year. Even at the apex of their season, NLCs do not encompass the entirety of Earth's surface, instead barely reaching the mid-latitudes. The duration of the crossing is still expected to be $10^4 - 10^6$ years, depending on the size of the GMC and the speed we travel through it.
However, we will show that other important climate effects appear. This paper is organized as follows: in the next section we describe the atmospheric model. In section \ref{sec:NLCs} we explore the properties of dense NLCs, including radiative transfer simulations. In section \ref{sec:ozone} we examine mesospheric ozone depletion. Finally, we summarize and give concluding remarks in section \ref{sec:discuss}.

%%%%%%%%%%%%%%%%%%%%%%%%%%%%%%%%%%%%%%%%%%%%%%%%%%%%%%%%%%%%%%%%%%%%%%%%%%%
\section{Model description}
\label{sec:model}

The atmospheric model we use is the two-dimensional chemistry-dynamics model developed at NASA Goddard Space Flight Center \cite{douglass_comparison_1989, jackman_effect_1990, considine_effects_1994, jackman_past_1996, fleming_simulation_1999, jackman_northern_2001, thomas_terrestrial_2023}.
This model has 76 altitude bins, 45 latitude bins, and traces 80 chemical species as well as temperature and wind dynamics, all zonally-averaged. We use pre-industrial boundary conditions on the surface, in which anthropogenic compounds such as CFCs (chlorofluorocarbons) and HCFCs (hydrochlorofluorocarbons) have zero surface flux. Greenhouse gases are set to pre-industrial concentrations as well.
While CO$_2$ levels have fluctuated between 300-400 ppmv in the last 5 Myr \cite{seki_alkenone_2010}, the exact timing when the solar system crossed the LxCC is unclear \cite{opher_possible_2024}, so we adopt pre-industrial value of 280 ppmv. The higher CO$_2$ mixing ratio in the paleoclimate may actually lower the temperature near the mesopause, promoting NLC growth \protect \cite{roble_how_1989}.

The concentration of interstellar hydrogen entering the upper atmosphere is a function of the interstellar flux,
\begin{equation}
    \phi = n v,
\end{equation}
where $n$ is the number density of interstellar hydrogen and $v$ is the interstellar cloud velocity. We relate the incoming hydrogen flux to water mixing ratio outlined by \citeA{mckay_consequences_1978} to obtain a value for the mixing ratio of water in the upper mesosphere/lower thermosphere (at 90 km in our model). This mixing ratio is overwritten every timestep to enact the continual interstellar hydrogen flux.

We perform three main simulations: a control run without any water input, one for the LxCC and one for the Local Bubble, as the latter two have different hydrogen fluxes and therefore different water concentrations. For the LxCC, we use a density and velocity of 6200 cm$^{-3}$ and 40 km/s, yielding an H$_2$O mixing ratio of 566 ppm. Similarly, for the Local Bubble, the density and velocity of 1200 cm$^{-3}$ and 70 km/s yield a mixing ratio of 113 ppm. These parameters come from the MHD simulations of \citeA{opher_possible_2024, opher_passage}. Simulations are run for 24 years to ensure they come to equilibrium.

%%%%%%%%%%%%%%%%%%%%%%%%%%%%%%%%%%%%%%%%%%%%%%%%%%%%%%%%%%%%%%%%%%%%%%%%%%%
\section{Noctilucent cloud coverage}
\label{sec:NLCs}

Noctilucent clouds (NLCs, also called polar mesospheric clouds, PMCs) are the highest altitude clouds in the atmosphere and form just below the polar mesopause in the summer when the atmosphere reaches its coldest temperature ($\sim$140 K). By enhancing the water concentration in this region, NLCs should become denser and have longer seasonal durations.

In this atmospheric model, NLCs are implemented as a transition from water vapor to ice when the air becomes saturated with water. This process depends heavily on having a high water concentration and low temperature. This calculation is performed with a simple model \cite{hervig_relationships_2009} without regard to condensation nuclei or other details of ice grain formation. Any ice transported out of the NLC region is converted back into water vapor. While some temperature differences between the northern and southern mesosphere on the order of 3-9 K are expected \cite{hervig_inter-hemispheric_2013, siskind_hemispheric_2003}, the southern mesosphere in our model does not reach temperatures low enough to form NLCs (only 161 K at its lowest), while the northern hemisphere matches observations more accurately. As such, we only consider NLCs in the northern hemisphere.

\begin{figure}
\centering
\noindent\includegraphics[width=0.95\textwidth]{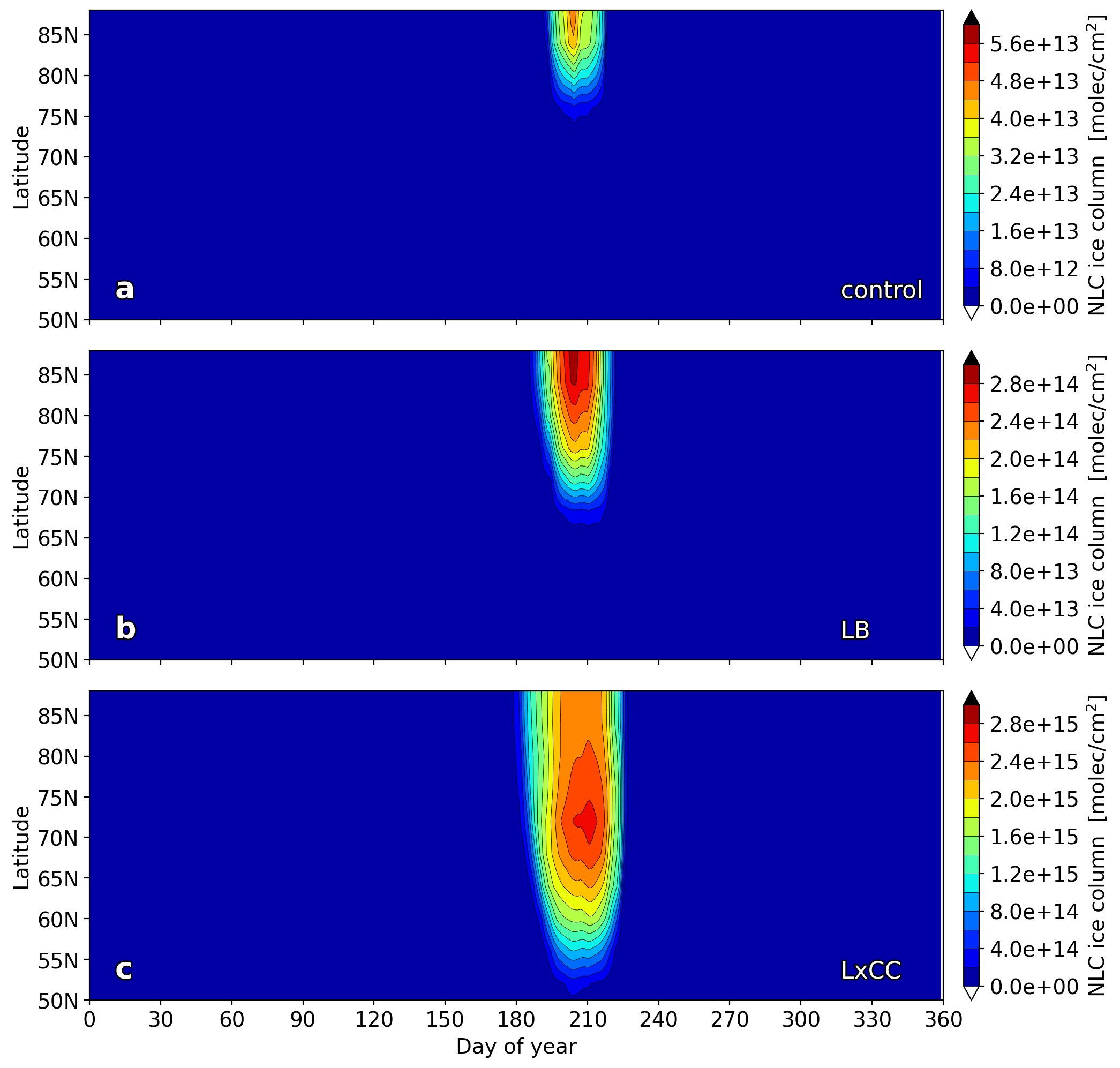}
\caption{Solid H$_2$O (ice) column density during the final year in the control (a), Local Bubble (b), and LxCC (c) simulations. Note that the color scales in (b) and (c) are 5 and 50 times larger than in (a), respectively. Only altitudes greater than 60 km are included in order to isolate NLCs.}
\label{fig:NLCcolumn}
\end{figure}

To probe the NLC thickness and density, the ice column density is calculated. Fig. \ref{fig:NLCcolumn} shows the mesospheric ice column density for NLCs over the course of the final simulation year in the control, Local Bubble, and LxCC simulations. The peak column density in the control is $4.65 \times 10^{13}$ molec cm$^{-2}$. The peak values in the Local Bubble and LxCC simulations are $2.89 \times 10^{14}$ molec cm$^{-2}$ and $2.69 \times 10^{15}$ molec cm$^{-2}$ respectively, significantly higher than that of the control. The seasonal duration is similarly enhanced, increasing from 25 days to 34 (Local Bubble) and 45 (LxCC) days. The lowest latitude the NLCs reach also extends from 78\degree N down to 67\degree N (Local Bubble) and 51\degree N (LxCC). In all aspects, NLCs formed as Earth crosses an interstellar cloud are stronger and more prevalent than the ones seen today.

\citeA{mckay_consequences_1978} hypothesized that an interstellar cloud would cause global NLC coverage that decreases warming and, over sustained times, may incur a runaway effect that triggers an ice age. In contrast, the simulations presented here show a significant increase in NLC coverage, but it is not nearly global in scale. Despite the higher water concentration, the mesopause temperature still only falls low enough to form NLCs in the summer polar regions. To test this, we performed an LxCC simulation in which the temperature when calculating NLC ice concentration was artificially lowered by 20 K. While the maximum ice concentration of the NLCs grew by an order of magnitude, the latitudinal extent encompasses the poles and mid-latitudes down to 30\degree N. Furthermore, the NLC season increased to 90 days. This extent and duration would be note-worthy, but is still far from the permanent global coverage suggested by \citeA{mckay_consequences_1978}.

Nevertheless, \citeA{mckay_consequences_1978} are qualitatively correct in mentioning that denser clouds block more incoming sunlight. To investigate the attenuation of sunlight due to NLCs more quantitatively, we employed offline radiative transfer simulations using the UVSPEC model of the libRadtran package \cite{emde_libradtran_2016, mayer_technical_2005}. 
Following the same approach as \citeA{lange_colour_2022}, we assumed spherical NLC particles, and we used the Mie code of libRadtran package to calculate the single scattering properties for ice grains of 20 nm radius. These optical properties were used as input in UVSPEC, and simulations were performed for dense NLCs 1 km thick at an altitude of 80 km for a range of ice water content (IWC) values. Table \ref{tab:rad} shows different calculated solar irradiance components, the global and diffuse upwards irradiances, at the surface of Earth and at the top of atmosphere (TOA), respectively.
These are instantaneous results located at latitude 85\degree N and longitude 10\degree E, with a solar zenith angle (SZA) of 60\degree. The overall effect is slight except for very high IWC. Over the course of one day as the SZA progresses, an NLC with an IWC of 0.01 g/m$^3$ alters the incoming solar irradiance at the surface by a relatively steady 7-8\%. We note that to investigate the complete climate effect, a more detailed climate model is needed that couples NLC ice grain density to a radiation transfer scheme.

\begin{table*}[htb]
    \centering
    \caption{Results of instantaneous radiative transfer simulations for NLCs of varying optical depths}
    \label{tab:rad}
    \begin{tabular}{ccccc}
    \hline
    Ice Water & Optical & Global irradiance & Diffuse upwards irradiance & Net irradiance\\
    Content (g/m$^3$) & depth & at surface (W/m$^2$) & at TOA (W/m$^2$) & at TOA (W/m$^2$) \\
    \hline
    0 & 0 & 445.69 & 42.9 & 563.76 \\
    $2\times 10^{-14}$ & 0 & 445.69 & 42.9 & 563.75 \\
    0.0001 & 0.001 & 445.29 & 43.4 & 563.20 \\
    0.001 & 0.01 & 441.84 & 48.2 & 558.41 \\
    0.01 & 0.1 & 412.74 & 83.4 & 523.3 \\
    \hline
 \end{tabular}
\end{table*}

Interstellar clouds contain dust grains that would be transported directly to Earth's atmosphere. Due to the high speed of interstellar dust, these grains would at least partially vaporize in the mesosphere upon impact \cite{vondrak_chemical_2008}, leading to a higher concentration of condensation nuclei (CN) for the NLCs to condense upon. The dust flux may be a factor of up to 100 times greater than the present value, depending on the density of the interstellar cloud \cite{pavlov_passing_2005}. The relation between CN and the radiative transfer properties of NLCs is still unclear \cite{megner_minimal_2011, wilms_nucleation_2016}. It is possible that more CN produce smaller grains that interact with light less strongly, leading to weaker radiative effects. In this case, NLCs would not cause a large surface temperature change, as suggested by \citeA{mckay_consequences_1978}.

\begin{figure}
\centering
\noindent\includegraphics[width=0.95\textwidth]{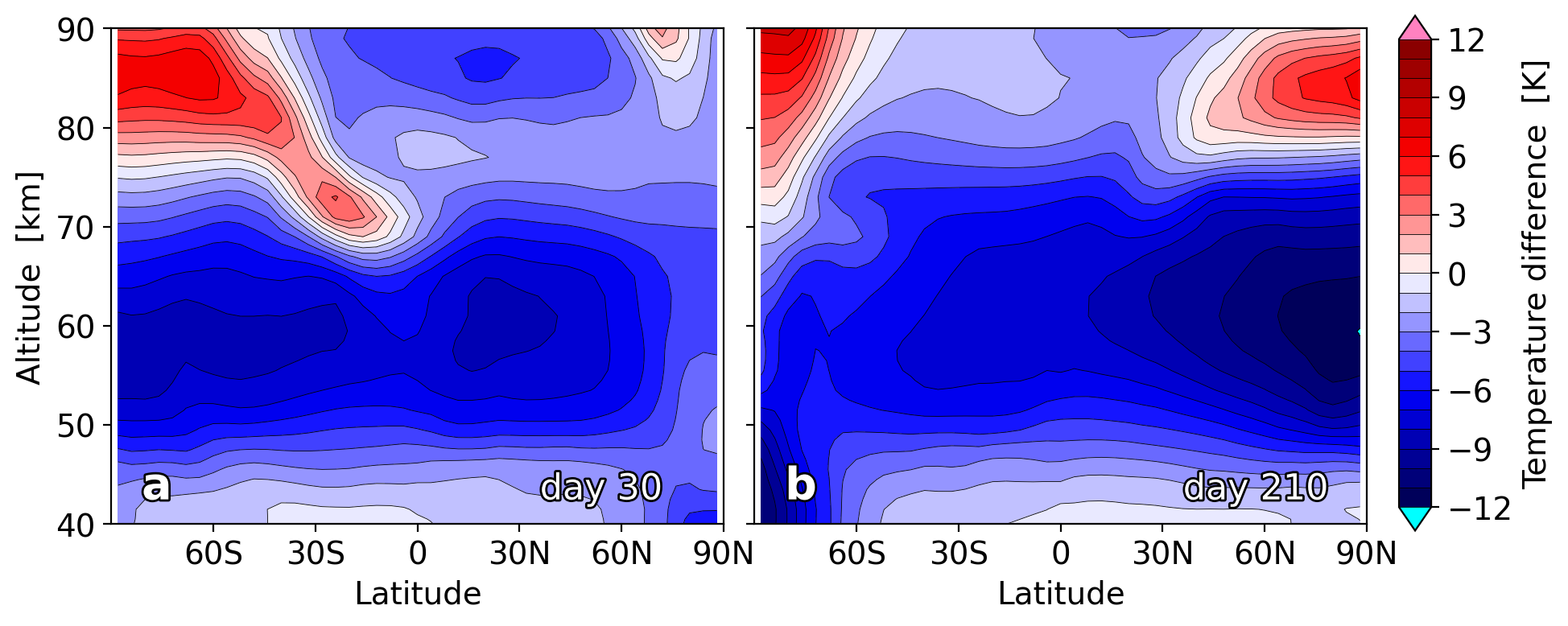}
\caption{Altitude-latitude plots of the difference in temperature on days 30 (a) and 210 (b) between the LxCC and the control run. The maximum temperature decrease on day 210 reaches 12 K.}
\label{fig:temp}
\end{figure}

The temperature throughout the mesopause is significantly affected by the introduction of additional water. This is most apparent in the vernal and autumnal mesosphere, outside of the NLC season. Unlike NLCs, the extent is global. Fig. \ref{fig:temp} shows a snapshot of the temperature difference between the control and LxCC simulations on days 30 and 210, highlighting the more extreme changes. A global cooling is apparent between 45-70 km across all latitudes for most of the year. In the most extreme, the temperature drops by 12 K.

%%%%%%%%%%%%%%%%%%%%%%%%%%%%%%%%%%%%%%%%%%%%%%%%%%%%%%%%%%%%%%%%%%%%%%%%%%%
\section{Ozone depletion}
\label{sec:ozone}

The introduction of high-altitude water participates in a set of catalytic chemical reactions that deplete ozone. Most ozone exists in the stratosphere, with higher altitude ozone through the mesosphere dropping off exponentially. When crossing an interstellar cloud, \citeA{yeghikyan_terrestrial_2004} found a 50-80\% decrease in ozone near 80 km.

The conversion of water into HO$_x$ happens via photolysis as
\begin{equation}
    H_2O + h\nu \rightarrow OH + H,
\end{equation}
wheren $h\nu$ is a photon less than 240 nm. HO$_x$ compounds (H, OH, HO$_2$) have long been known to deplete ozone via the catalytic process \cite<e.g.,>{bates_photochemistry_1950}
\begin{equation}
    \begin{array}{l}
         OH + O_3 \rightarrow HO_2 + O_2 \\
         HO_2 + O \rightarrow OH + O_2,
    \end{array}
\end{equation}
in which ozone is converted to O$_2$. Therefore, the ozone depletion is expected to occur concomitantly with the increased abundance of HO$_x$, which is observed in the model.

\begin{figure}
\centering
\noindent\includegraphics[width=\textwidth]{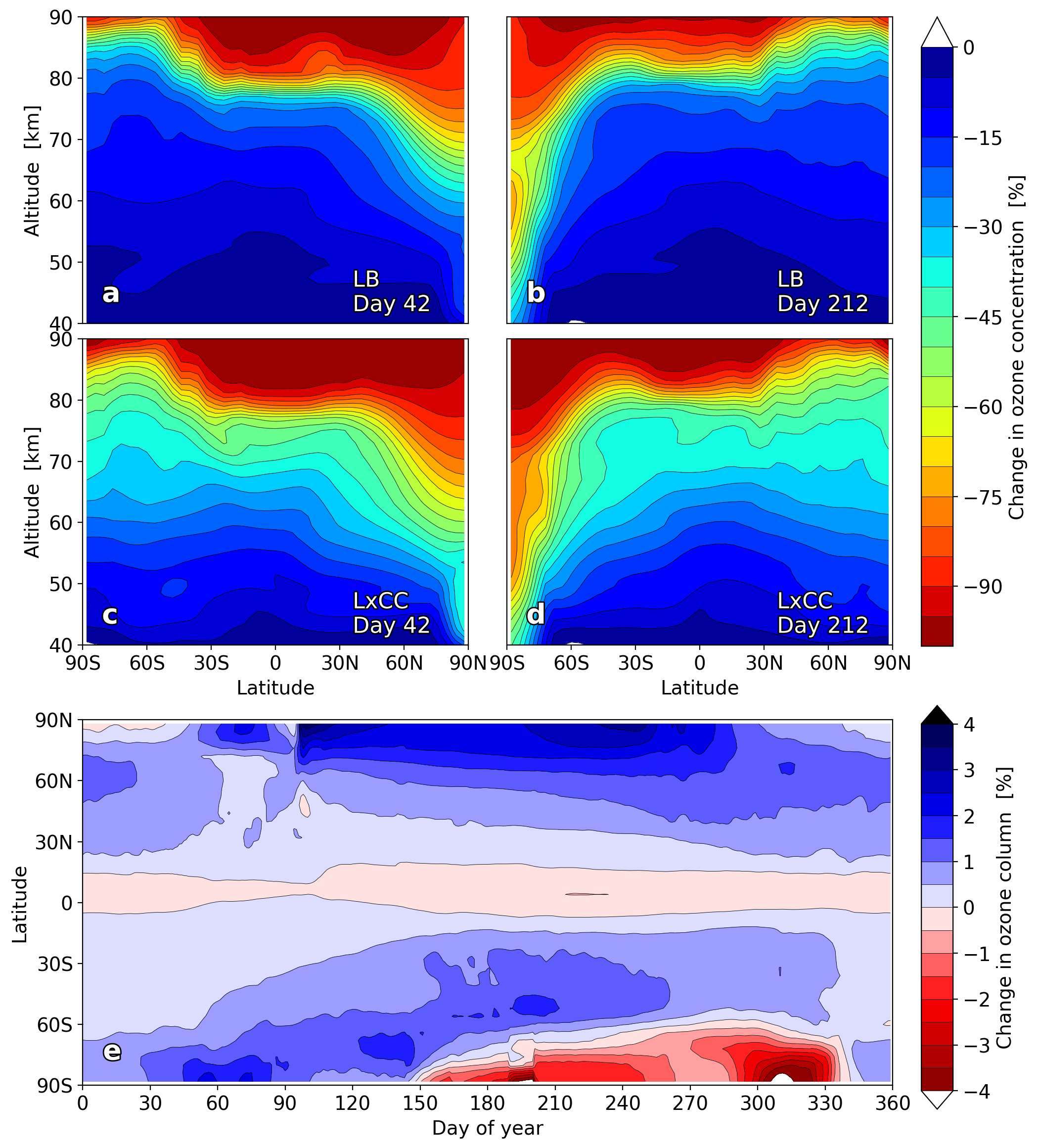}
\caption{Altitude-latitude plots of the percentage change in ozone concentration between the control case and that of the LxCC (a, b) and that of the Local Bubble (LB) wall (c, d). Panels (a) and (c) are taken on day 42 of the final year, and panels (b) and (d) are taken on day 212. Panel (e) is the percentage change in the total column density of ozone in the LxCC simulation.}
\label{fig:ozone}
\end{figure}

In Fig. \ref{fig:ozone}(a-d), the decrease in ozone is shown for two opposite days of the year, highlighting seasonal effects. Note that this is a percentage depletion, not the bulk amount (which varies by orders of magnitude). As expected, most of the ozone depletion occurs in the upper mesosphere. Near the solstices, more significant depletion occurs near the poles. This happens because the pole is pointed away from the Sun, blocking the UV light that makes ozone and amplifying the depletion. Effects of the abundance of interstellar hydrogen are also seen: the simulation showing the passage through the Local Bubble wall has the same broad features as that of the LxCC, but at a diminished capacity. This is a direct result of the lower H$_2$O density in the cloud at the wall of the Local Bubble.

It should be noted that, although the concentration of ozone changes dramatically in the upper mesosphere, it accounts for a small portion of the total ozone column, which is located primarily in the stratosphere. In fact, due to the lower ozone column in the mesosphere, UV light penetrates more deeply into the atmosphere, generating more ozone in the stratosphere. This is apparent in Fig. \ref{fig:ozone}(e), which shows that the majority of the change in the total ozone column is a slight increase. The only location of an overall decrease in the ozone column is during the winter and spring in the southern polar region. The sharp changes seen in the north around day 100 correspond to polar spring, in which the sun rapidly comes above the horizon and has a significant and sudden photochemical effect \cite[e.g.]{thomas_gamma-ray_2005}. The maximum change in the ozone column across the whole Earth is $\pm$4\%, both of which happen at the poles. The ozone column around the equatorial region, however, generally changes by less than 1\%.

Overall, this model predicts a drastic change in the ozone concentration in the upper mesosphere. However, this accounts for a small fraction of the ozone column, which actually increases. While the decrease in ozone was expected and also predicted by \citeA{yeghikyan_terrestrial_2004}, they did not account for any increased stratospheric ozone. As mentioned earlier, this ozone analysis can (and should) be further refined by including the effects of cosmic rays, which will act to deplete stratospheric ozone.

%%%%%%%%%%%%%%%%%%%%%%%%%%%%%%%%%%%%%%%%%%%%%%%%%%%%%%%%%%%%%%%%%%%%%%%%%%%
\section{Discussion and conclusions}
\label{sec:discuss}

We have modeled how Earth's mesosphere responded to dense interstellar clouds during two crossings in the past 10 Myr. The largest effects are the increased NLC coverage and the depletion of mesospheric ozone.

In contrast to \citeA{mckay_consequences_1978}, these results do not show global NLC formation, but instead maintain NLC confinement to the summer polar mesosphere. The low temperature required to form NLCs is too localized to this region to allow global NLC formation.
\citeA{mckay_consequences_1978} qualitatively suggested that the lack of mesospheric ozone would result in a colder mesopause and near-global NLC coverage. Applying our more complete model, not only does the mesopause temperature not globally decrease, it often increases. We see a temperature decrease in the lower- and mid-mesosphere, but not the upper region.
If the NLCs are dense enough, their direct radiative effect through blocking part of the incoming solar radiation may lead, in the long term, to a cooling of the Earth's system. At the same time, these cold high-altitude clouds can absorb a considerable amount of the outgoing thermal radiation. The extent and magnitude of these effects on the radiative budget, while outside the scope of this mesosphere study, deserve further attention, as they may have important implications on climate by altering planetary albedo and atmospheric dynamics.

Effects on ozone are largely similar results as those found by \citeA{yeghikyan_terrestrial_2004} in the mesosphere: high depletion of ozone in the upper mesosphere that trails off to less depletion further down. However, ozone depletion in the mesosphere does not mean a lower ozone column, as there is a resulting increase in the total ozone column, mostly in the polar regions. This happens due to the lower penetration depth of UV that generates ozone in the stratosphere. The total column is increased by at most 4\%. Given that this study is primarily focused on the mesosphere, the overall climatological effects of stratospheric ozone are left for a future investigation.

Additionally, exposing the Earth to the interstellar medium means the removal of the heliospheric shield that protects us from Galactic cosmic rays \cite<GCRs,>{opher_possible_2024, pavlov_catastrophic_2005}. These additional CRs have energies on the order of $\sim$ 1 GeV. However, CRs that contribute to atmospheric chemical production of NO$_x$ generally have energies $>$10 GeV. Earth's magnetic field applies some shielding and introduces latitudinal variations \cite{smart_magnetospheric_2000}, but this effect is left for a future study.

Because our results show that NLCs are neither permanent nor global in scope, they are less likely to cause an ice age as previous thought. As a result, our solar system's passages through dense interstellar clouds may be more difficult to detect throughout Earth's history than previously thought. While effects on the mesosphere are slight, additional potential effects on other parts of the atmosphere should be examined with a more detailed model such as WACCM that includes the full effects of higher-altitude chemistry and dynamics. Radiative effects of NLCs vary over the course of days and years, and so deserve further investigation.
Finally, further study is needed to identify whether the geological record could still carry the memory of such crossings.

%%% End of body of article

%%%%%%%%%%%%%%%%%%%%%%%%%%%%%%%%%%%%%%%%%%%%%%%
\section{Open Research}
Data is freely available at \citeA{miller_dataset}.
% The NASA GSFC atmosphere model outputs in netCDF format and the Jupyter Notebook used for plotting will be freely available through Zenodo at \citeA{miller_dataset}.
% \url{https://zenodo.org/doi/10.5281/zenodo.10658966}.

%%%%%%%%%%%%%%%%%%%%%%%%%%%%%%%%%%%%%%%%%%%%%%%
\acknowledgments
We graciously thank Mark Hervig for providing his NLC code.
This work is supported by NASA grant 18-DRIVE18\_2-0029 as part of the NASA/DRIVE program entitled ``Our Heliospheric Shield'', 80NSSC22M0164, \url{https://shielddrivecenter.com}.

%%%%%%%%%%%%%%%%%%%%%%%%%%%%%%%%%%%%%%%%%%%%%%%

\bibliography{bibliography}

\end{document}